\documentclass[11pt]{article}
\usepackage{mathrsfs}
\usepackage{amsmath}
\usepackage{amssymb}
\usepackage{amsthm}
\usepackage{cases}
\usepackage{aliascnt}
\usepackage{hyperref}
 \allowdisplaybreaks[4]
\theoremstyle{definition}
\newtheorem{theorem}{Theorem}[section]
\newcommand{\theoremref}[1]{Theorem~\ref{#1}}
\newaliascnt{lemma}{theorem}
\newtheorem{lemma}[lemma]{Lemma}
\aliascntresetthe{lemma}

\newcommand{\lemmaref}[1]{Lemma~\ref{#1}}
\newaliascnt{corollary}{theorem}
\newtheorem{corollary}[corollary]{Corollary}
\aliascntresetthe{corollary}

\makeatother       

\makeatletter      
\@addtoreset{equation}{section}
\makeatother       

\title{ \Huge  Precise constraints on a $\tau$ function in 2D quantum
gravity}
\author{  Liu Shaowei\\
\scriptsize College of  Mathematics and statistics, Southwest
University,\\ \scriptsize Beibei District,Chongqing, 400715, P.R. China \\
 \scriptsize Email: swliu001@swu.edu.cn}
 \vspace{1cm}
\date{}
\begin{document}
\maketitle
\begin{center}
\begin{center}{\bf Abstract}\end{center}
\vspace{1cm}
\begin{minipage}{12cm}
{\quad For an arbitrary $p$, propose a new and computable method
which can determine the values of unknown constants in
constraints on a tau function which satisfies both the p-reduced
KP hierarchy and the sting equation. All the constants do not
equal $0$, unlike what people usually think of. With these
values, obtain the precise algebra that the constraints compose.
This algebra includes none of $\{t_{mp}\}$ and also includes the
Virasoro algebra as a subalgebra.}
 \end{minipage}
\end{center}

\vspace{2cm}
\newpage
\section*{\S 1.\ \ Introduction}
\setcounter{section}{1}

\quad  Quantum gravity is an interesting object in the current
research of mathematical physics. In 2D quantum gravity,
Kontsevich\cite{k} proved Witten's conjecture\cite{w2} that two
different approaches to 2D quantum gravity coincide. That is, a
partition function for the intersection theory of moduli space is
the logarithm of some $\tau$ function\cite{jim} which satisfies
the string equation and the 2-reduced KP hierarchy. Meanwhile,
using Kontsevich's matrix integral representation of the
partition function, Witten\cite{w1} showed the exponent of the
partition function is a vacuum vector for the Virasoro algebra.
Together with the conclusion\cite{dvv,fkn} that a $\tau$ function
which satisfies the string equation and the 2-reduced KP
hierarchy is equivalent to a vacuum vector for the Virasoro
algebra, he also obtained the equivalence of the two approaches.
Since integrable system has close connection with the string
theory and the intersection theory\cite{mo,di}, this conclusion
has been wildly researched with various methods in this
field\cite{liu1,liu2,dk4,dk3,am1}. An interesting problem is to
extend the conclusion in \cite{dvv,fkn} from 2 to an arbitrary
$p$, which is to obtain the equivalence between a $\tau$ function
constrained by the string equation and the p-reduced KP hierarchy
and a vacuum vector of some algebra which include the Virasoro
algebra as a subalgebra. When $p=3$, Goeree\cite{go} showed that
it is true. And the case for bigger p had also been researched in
\cite{fkn,mo,am1}.

In order to obtain the above equivalence for an arbitrary $p$, it
need to obtain the precise constraints which the KP hierarchy and
the string equation impose on tau functions. When we calculate
them, it creates a lots of constants in the obtained constraints
whose values are unknown. When $p=2$, the constrains constitute
the Virasoro algebra and we could use the commutation relations
of the Virasoro algebra to calculate the values of the constants.
But when $p\geq 4$, although there are some classical conclusions
about $W$ algebra\cite{fkn,am1}, it is so hard to calculate the
commutation relations of the constraints that the constants in
the higher order constraints are still unknown. As far as we have
known, there is not an effective computable method to determine
them when $p\geq 4$. Due to the uncertainty of the constants, we
also could not obtain the precise algebraic structure of these
constraints. In this study, for an arbitrary $p$, we propose a
new computable method which can determine values of the
constants. It is a recursive process and we can directly
calculate them step by step. It is usually that assign the value
of $0$ to all the constants; but here, by this method, we know
that all of them are not equal $0$. And the none zero constants
are closely related with the centers of the algebra that the
constrains constitute. When $p=2$, our conclusion coincide with
the current conclusion, that is, the constants determined through
our method being the same as those determined through commutation
relations of Virasoro. Consequently, with these values we obtain
the precise constraints. And further we obtain the precise
algebra which the constraints constitute. The algebra does not
include the redundant variables of $t_{mp}$, and we find the
connection between the algebra and the $W_{1+\infty}$ algebra
which include all the variables of $t_{mp}$. Based on this
connection, we can calculate the commutation relations of one
algebra from those of the other algebra. And the calculation is
much simpler than a straightforward calculation. Furthermore, the
obtained algebra include the Virasoro algebra as a subalgebra. So
the above conclusion in \cite{dvv,fkn} is also included in our
conclusion. In addition, we mainly use the tool of
pseudo-differential operators to prove the conclusions, which is
introduced by Dickey\cite{dk2} and greatly simplifies the proof.

The organization of the paper is as follows. In section 2, for
self-contained we give a brief description of the KP hierarchy.
In section 3, we prove the connection between  $W_{1+\infty}$
algebra and the algebra of $\bar{W}=\{W_{n}^{(m)}|_{t_{mp}=0}\}$.
In section 4, we show the approach to calculate unknown
constants, which is our main theorem. Meanwhile, we give some
examples for our theorems. Section 5 is devoted to conclusions.
\section*{\S 2.  KP hierarchy }
\setcounter{section}{2}

\quad To be self-contained, we give a brief introduction to the
KP hierarchy  based on a detailed research in \cite{dk2}.

Let $F$ be an associative ring of functions which include
infinite time variables $t_i\in\mathbb{R}$:
\[F=\left\{f(t)=f(t_1,t_2,\cdots,t_j,\cdots) ; t_i\in\mathbb{R}
\right\}.
\]
Denote $\partial_{t_1}$ by $\partial$, which is the common
differential operator on the first variable $t_1$. Its actions on
$f(t)$ are
\begin{equation}
\partial
f(t)=\partial_{t_1}f(t); \quad\quad\partial\circ
f(t)=f(t)\partial+\partial_{t_1}f(t).
\end{equation}
Here the symbol $``\circ"$ denote the multiplication between
operators. If we consider a function $f(t)$ as a operator whose
action on $g(t)\in F$ is $f(t)g(t)$, we can infer the following
identity about multiplication of function operators and
differential operators. That is for any $j\in\mathbb{Z}$
\begin{equation}
\partial^j\circ
f=\sum^{\infty}_{i=0}\binom{j}{i}(\partial^i
f)\partial^{j-i},\hspace{.3cm}
\binom{j}{i}=\frac{j(j-1)\cdots(j-i+1)}{i!},\quad
j\in\mathbb{Z}.\label{vvv21}
\end{equation}
If the function operators are located on the the left-hand side
we omit $``\circ"$. So with (\ref{vvv21}) we could obtain an
associative ring $F (\partial)$ of formal pseudo differential
operators, which includes two operations $``+"$ and $``\circ"$:
\[F(\partial)=\left\{R=\sum_{j=-\infty}^d f_j(t)\partial^j, f_j(t)\in
F\right\}.
\]
The ring $F(\partial)$ includes two subrings which are
$F_+(\partial)=\{R_+=\sum_{j=0}^d f_j(t)\partial^j\}$, the ring
of deference operators and
$F_-(\partial)=\{R_-=\sum_{j=-\infty}^{-1} f_j(t)\partial^j\}$,
the the ring of Volterra operators.

Let $L$ be a general first order pseudo differential operator:
\begin{equation}
L=\partial+ \sum_{j=1}^{\infty} f_j(t)\partial^{-j}.
\end{equation}
The KP-hierarchy\cite{dk2} can be expressed as
\begin{equation}
\frac{\partial L}{\partial t_i}=[(L^i)_+, L].\label{vvv27}
\end{equation}
Comparing the powers of $\partial$ on both sides, we can obtain
the family of equations in functions $f_j(t)$. From $i=2$ and
$i=3$, we obtain
\begin{equation}
3\partial_{t_2}^2f=\partial_{t_1}(4\partial_{t_3}f-\partial_{t_1}^3f-6f\partial_{t_1}f),
\end{equation}
which is the famous Kadomtsev-Petviashvili equation.

Define the dressing operator
\[\Phi(t)=1+\sum^\infty_{j=1}w_j(t)\partial^{-j},
\]
which satisfies
\begin{equation}
L=\Phi(t)\circ\partial\circ \Phi(t)^{-1}.
\end{equation}
Then the Sato equation in the operator $\Phi(t)$
\begin{equation}
\frac{\partial \Phi(t)}{\partial t_i}=-(L^i)_-\circ
\Phi(t)\label{sato}
\end{equation}
is equivalent to the KP equation (\ref{vvv27}). The KP hierarchy
can also be expressed as the following equations equivalently:
\begin{equation}
L^kw=z^kw\quad\text{and}\quad \partial_{t_m}w=L_+^mw.
\end{equation}
Using the dressing operator $\Phi$, we can give a kind of form
solutions for the above equations which are called Baker or wave
functions $w(t,z)$ and adjoint Baker or wave functions
$w^*(t,z)$:
\begin{align}
w(t,z)=\Phi(t)exp(\sum^\infty_{i=1}t_i z^i)
=\bigg(1+\frac{w_1(t)}{z}+\frac{w_2(t)}{z^2}+\cdots\bigg)\exp(\sum_{i=1}^{\infty}t_iz^i)
\end{align}
and
\begin{align}
w^*(t,z)=(\Phi^{-1}(t))^* exp(\sum^\infty_{i=1}-t_i z^i)
=\bigg(1+\frac{w_1^*(t)}{z}+\frac{w_2^*(t)}{z^2}+\cdots\bigg)\exp(\sum_{i=1}^{\infty}t_iz^i).
\end{align}

a KP hierarchy is equivalent to a single function, the tau
function $\tau(t)$, which means that all functions $w_i(t)$ in
the dressing operator $\Phi(t)$ can be generated by a single
function, the $\tau$ function. That is
\begin{equation}
w(t,z)=\frac{\tau(t-[z^{-1}])}{\tau(t)} exp(\sum^\infty_{i=1}t_i
z^i)\label{24}
\end{equation}
and
\begin{equation}
w^*(t,z)=\frac{\tau(t+[z^{-1}])}{\tau(t)} exp(\sum^\infty_{i=1}-t_i
z^i),\label{25}
\end{equation}
where $[z]=(z,\frac{z^2}{2},\frac{z^3}{3},\cdots)$. And further
all function $f_i(t)$ in the operator $L$ can be generated by the
tau function too. Here introduce the $G(z)$ operator, whose
action on functions is $G(z)f(t)=f(t-[z^{-1}])$.

There are a family of symmetries for KP hierarchy which are
called the additional symmetries. We denote the infinitesimal
operators for the symmetries by $\partial_{ml}^*$. Their actions
on on $\Phi(t)$ are as follow\footnote{Here for convenient the
symbol has a slight different from \cite{dk2}.}
\begin{equation}
\partial^*_{ml}\Phi(t)=-(M^mL^l)_-\circ \Phi(t).\label{radd}
\end{equation}
where
\begin{equation}
M=\Phi(t)\circ\Gamma\circ \Phi(t)^{-1},
\end{equation}
and $\Gamma=\sum_{i=1}^\infty it_i\partial^{i-1}$.

There are another family of symmetries for the KP hierarchy.
Their infinitesimal operators are called vertex operators, which
are defined as follow
\begin{equation}
X(\lambda,\mu)=:\exp\sum_{i=-\infty}^\infty(\frac{P_i}{i\lambda^i}-\frac{P_i}{i\mu^i}):
\end{equation}
where
\begin{eqnarray}
  P_i=
  \begin{cases}
    \partial_i,\quad &i>0\\ \\
    |i|t_{|i|},\quad &i\leq 0
  \end{cases}\notag
\end{eqnarray}
Talyor expand the $X(\lambda,\mu)$ in $\mu$ at the point of
$\lambda$, we have
\begin{equation}
X(\lambda,\mu)=\sum_{m=0}^\infty
\frac{(\mu-\lambda)^m}{m!}\sum_{n=-\infty}^\infty\lambda^{-n-m}W_n^{(m)},
\end{equation}
where
\[\sum_{n=-\infty}^\infty\lambda^{-n-m}W_n^{(m)}=\partial_\mu^m|_{\mu=\lambda} X(\lambda,\mu).
\]
By a straightforward computation, the first items of $W_n^{(i)}$
are as follow
\begin{align}
&W_n^{(0)}=\delta_{n,0},\\
&W_n^{(1)}=P_n,\\
&W_n^{(2)}=\sum_{i+j=n}:P_iP_j:-(n+1)P_n,\\
&W_n^{(3)}=\sum_{i+j+k=n}:P_iP_jP_k:-\frac{3}{2}(n+2)\sum_{i+j=n}:P_iP_j:+(n+1)(n+2)P_n,\\
&W_n^{(4)}=P_n^{(4)}-2(n+3)P_n^{(3)}+(2n^2+9n+11)P_n^{(2)}-(n+1)(n+2)(n+3)P_n\\
&W_n^{(5)}=P_n^{(5)}-\frac{5}{2}(n+4)P_n^{(4)}+(\frac{10}{3}n^2+20n+35)P_n^{(3)}\notag\\
&-(\frac{5}{2}n^3+20n^2+\frac{105}{2}n+50)P_n^{(2)}+(n+1)(n+2)(n+3)(n+4)P_n\\
&\cdots\cdots\notag
\end{align}
where
\begin{align}
&P_n^{(0)}=\delta_{n,0},\\
&P_n^{(1)}=P_i\\
&P_n^{(2)}=\sum_{i+j=n}:P_iP_j:\\
&P_n^{(3)}=\sum_{i+j+k=n}:P_iP_jP_k:\\
&P_n^{(4)}=\sum_{i+j+k+l=n}:P_iP_jP_kP_l:-\sum_{i+j=n}:ijP_iP_j:\\
&P_n^{(5)}=\sum_{i+j+k+l+m=n}:P_iP_jP_kP_lP_m:-5\sum_{i+j+k=n}:ijP_iP_jP_k:\\
&\cdots\cdots\notag
\end{align}

The two families of symmetries are equivalent. This fact is
represented by the ASvM formula \cite{dk3,asm2,asm3}
\begin{equation}
\partial^*_{m,l+m}\tau(t)=\frac{W_l^{{(m+1)}}\cdot
\tau(t)}{m+1},\label{r2}
\end{equation}
which hold for $m\geq 0$ and for all $l$.

In integrable system, using the additional symmetries, the string
equation \cite{dk2} can be represent as
\begin{equation}
[L^p,\frac{1}{p}(ML^{-p+1})_+]=1,
\end{equation}
or equivalently as
\begin{equation}
  \partial^*_{1,-p+1}L=0.\label{61}
\end{equation}

\section*{\S 3 Constraint equations and connection between two algebras}
\setcounter{section}{3} \setcounter{equation}{0}

\quad In this section, firstly we show equations that a tau
function, which is under constraints of both the string equation
and the p-reduced KP, satisfies. Similar results were obtained by
other method. Here we use the method of pseudo-differential
operators to obtain them, which simplifies the proof. Secondly,
we find the connection between the $W_{1+\infty}$ algebra which
includes the redundant variables of $\{t_{mp}\}$ and the algebra
of $\bar{W}=\{W_{n}^{(m)}|_{t_{mp}=0}\}$ which doesn't include
$\{t_{mp}\}$. This connection is used in the proof of
\autoref{constants} in next section. Furthermore, using this
connection, we could greatly simplify the calculations about
commutation relations of $\bar{W}$.

Now, to obtain the constraint equations for a corresponding
$\tau$ function, we first show actions of additional symmetries
on wave functions.

\begin{lemma}\label{lemma>a}
 For any integer $ m\geq 0$ and $l\in
\mathbb{Z}$,
\begin{equation}
  \partial^*_{m,m+l}w(t,z)=\Bigg(\Big(G(z)-1\Big)\dfrac{\dfrac{W_l^{(m+1)}}{m+1}\cdot\tau(t)}{\tau(t)}\Bigg)\cdot w(t,z).
\end{equation}
\end{lemma}
\noindent{\sl {\bf   Proof:}} By (\ref{24}) and (\ref{r2}),
\begin{align}
&\partial^*_{m,m+l}w(t,z)\\=&\partial^*_{m,m+l}\bigg(\frac{G(z)\tau(t)}{\tau(t)}\exp\sum_{i=1}^\infty
(t_iz^i)\bigg)\notag\\=&\frac{\tau(t)\cdot
G(z)(\partial^*_{m,m+l}\tau(t))-G(z)\tau(t)\cdot\partial^*_{m,m+l}\tau(t)}{\tau^2(t)}\exp\sum_{i=1}^\infty
(t_iz^i)\notag\\
=&\Bigg(\Big(G(z)-1\Big)\dfrac{\dfrac{W_l^{(m+1)}}{m+1}\cdot\tau(t)}{\tau(t)}\Bigg)\cdot
w(t,z).\notag
\end{align}
$\hfill{\Box}$

Now we consider the string equation. From \cite{am1}, we know
that if the operator $L$ satisfies the p-reduced KP and the
string equation, then
\begin{align}
(M^jL^{kp+j})_-&=\prod_{r=0}^{j-1}(\frac{p-1}{2}-r)L^{-p}\quad&&\text{when}\quad
k=-1;\quad j=1,2,\cdots\\
&=0\quad &&\text{when}\quad k=0,1,2,\cdots;\quad j=1,2,\cdots
\end{align}
Substitute the above into the definition of the additional
symmetry, we have
\begin{align}
\partial^*_{j,kp+j}\Phi(t)&=-\prod_{r=0}^{j-1}(\frac{p-1}{2}-r)L^{-p}\circ\Phi(t)\quad
&&\text{when}\quad
k=-1;\quad j=1,2,\cdots\\
&=0\quad&&\text{when}\quad k=0,1,2,\cdots; \quad
j=1,2,\cdots\cdots
\end{align}
Substitute the above into the definition of the wave functions,
we have
\begin{align}
\partial^*_{j,kp+j}w(t,z)&=-\prod_{r=0}^{j-1}(\frac{p-1}{2}-r)z^{-p}\quad&&\text{when}\quad
k=-1;\quad j=1,2,\cdots\\
&=0\quad &&\text{when}\quad k=0,1,2,\cdots; \quad j=1,2,\cdots\label{68}
\end{align}
Then, using \lemmaref{lemma>a}, we could obtain the constraint
equations for the $\tau$ function which satisfies both the string
equation and the p-reduced KP.

\begin{corollary}\label{coro1}
If a $\tau$ function satisfies both the string equation and the
p-reduced KP, then it satisfies
\begin{align}
(G(z)-1)\dfrac{\dfrac{W_{kp}^{{(j+1)}}}{j+1}\cdot\tau(t)}{\tau(t)}&=-\prod_{r=0}^{j-1}(\frac{p-1}{2}-r)z^{-p}\quad&&\text{when}\quad
k=-1;\quad j=1,2,\quad\cdots\label{p4}\\
&=0\quad &&\text{when}\quad k=0,1,2,\cdots; \quad j=1,2,\cdots\label{69}
\end{align}
\end{corollary}
Now we consider the connection between the $W_{1+\infty}$ algebra
which includes the redundant variables of $\{t_{mp}\}$ and the
algebra of $\bar{W}$ which doesn't include $\{t_{mp}\}$.

Firstly, we give some notations here. From the expression of
$P_{kp}^{(i)}$, we know that every $P_{kp}^{(i)}$ is a summation
of items of normal product. For each $P_{kp}^{(i)}$, these items
can be divided into two categories; one includes all items that
comprise none of variables in $\{t_{mp}|m\in\mathbb{N}\}$ and the
other includes all items that comprise at least one variable in
$\{t_{mp}\}$. We represent the summation of all items in the
former category by $\bar{P}_{kp}^{(i)}$ and the summation of all
items in the latter category by $\hat{P}_{kp}^{(i)}$. So we have
$P_{kp}^{(i)}=\bar{P}_{kp}^{(i)}+\hat{P}_{kp}^{(i)}$ and
$\bar{P}_{kp}^{(i)}=P_{kp}^{(i)}|_{t_{mp}=0}$ for every
$P_{kp}^{(i)}$. Similarly we denote
$P_{kp}^{(i)}|_{\substack{t_{mp}=0\\ \partial_{mp}=0}}$ and the
other part by $\tilde{P}_{kp}^{(i)}$ and $\check{P}_{kp}^{(i)}$
respectively. And these notations are also applicable to
$\{W_{kp}^{(i)}\}$.

Now we give the connection between the algebra of $W_{kp}^{(i)}$
and the algebra of $\bar{W}_{kp}^{(i)}$.

\begin{lemma}\label{wa}
Expand $W_{kp}^{(n)}$ in $\{P_{-mp}|m\in\mathbb{N}\}$; then the
coefficient of $P_{-m_1p}\cdot P_{-m_2p}\cdot...\cdot P_{-m_ip}$
are linear combination of $\bar{W}_{(m_1+m_2+...+m_i+k)p}^{(l)},
l=1,...,n$. That is
\begin{align}
W_{kp}^{(n)}=\bar{W}_{kp}^{(n)}+\sum_{i=1}^{n}\sum_{m_1,m_2,...,m_i\in\mathbb{N}}P_{-m_1p}P_{-m_2p}\cdot...\cdot
P_{-m_ip}(constant\cdot
\bar{W}_{(m_1+m_2+...+m_i+k)p}^{(n-i)}\notag\\
+constant\cdot\bar{W}_{(m_1+m_2+...+m_i+k)p}^{(n-i-1)}+...+constant\cdot
\bar{W}_{(m_1+m_2+...+m_i+k)p}^{(0)}),\label{waa}
\end{align}
where the constants only depend on $p$ and especially the
constant in front of $P_{-m_ip}\cdot \bar{W}_{(m_i+k)p}^{(n-1)}$
equals $n$.
\end{lemma}
\noindent{\sl{\bf Proof:}} First we prove a more general
conclusion, the expansion of $\{W_{n}^{(m)}|n\in\mathbb{Z}\}$ in
$P_{-mp}$. Due to the normal product, we have
\begin{align}
&\sum_{n=-\infty}^\infty\lambda^{-n-m}W_n^{(m)}\label{wa3}\\
=&\partial_\mu^m|_{\mu=\lambda}:\exp\sum_{i=-\infty}^\infty(\frac{P_i}{i\lambda^i}-\frac{P_i}{i\mu^i}):\\
=&\partial_\mu^m|_{\mu=\lambda}\bigg(\exp\sum_{i=1}^{\infty}\frac{P_{-ip}\cdot\mu^{ip}}{ip}:\exp(\sum_{i=-\infty}^\infty\frac{P_i}{i\lambda^i}-\sum_{\substack{i=-\infty\\i\neq-kp}}^{\infty}\frac{P_i}{i\mu^i}):\bigg)\\
=&\sum_{l=0}^{m}C_m^l\cdot\partial_\mu^{l}|_{\mu=\lambda}
\bigg(\exp\sum_{i=1}^{\infty}\frac{P_{-ip}\cdot\mu^{ip}}{ip}\bigg)\cdot C_m^{m-l}\partial_\mu^{m-l}|_{\substack{\mu=\lambda\\t_{kp}=0}}\bigg(:\exp\sum_{i=-\infty}^\infty(\frac{P_i}{i\lambda^i}-\frac{P_i}{i\mu^i}):\bigg)\label{waa1}\\
=&\sum_{l=0}^{m}\bigg(C_m^l\sum_{i=0}^{l}\sum_{m_i\in\mathbb{N}}constant\cdot
(\prod_{i}P_{-m_ip})
\cdot\lambda^{(m_1+m_2+...m_i)p-l}\bigg)\bigg(C_m^{m-l}\sum_{n=-\infty}^\infty\lambda^{-n-(m-l)}\bar{W}_n^{(m-l)}\bigg).\label{wa4}
\end{align}
For any fixed $n$ and $m$, compare the coefficient of
$\lambda^{-n-m}$ on both sides. Then we have
\begin{align}
W_n^{(m)}&=\bar{W}_n^{(m)}+m\cdot\sum_{m_1\in\mathbb{N}}
P_{-m_1p}(\bar{W}_{m_1p+n}^{(m-1)}
+constant\cdot\bar{W}_{m_1p+n}^{(m-2)}+...+constant\cdot
\bar{W}_{m_1p+n}^{(0)})\\&+\sum_{i=2}^{n}\sum_{m_1,m_2,...,m_i\in\mathbb{N}}P_{-m_1p}P_{-m_2p}\cdot...\cdot
P_{-m_ip}(constant\cdot
\bar{W}_{(m_1+m_2+...+m_i)p+n}^{(m-i)}\notag\\
&+constant\cdot\bar{W}_{(m_1+m_2+...+m_i)p+n}^{(m-1-i)}+...+constant\cdot
\bar{W}_{(m_1+m_2+...+m_i)p+n}^{(0)})
\end{align}
Then, let $n=kp$ and we obtain that the coefficient of
$P_{-m_1p}\cdot P_{-m_2p}\cdot...\cdot P_{-m_ip}$ are linear
combination of $\bar{W}_{(m_1+m_2+...+m_i+k)p}^{(l)},
l=1,...,n-1$.

Meanwhile, the constants in (\ref{waa}) are linear combination of
the constants in the first bracket in (\ref{waa1}). Since the
latter are only depend on $p$ when $n=kp$, we obtain that the
constants in (\ref{waa}) are only depend on $p$.

$\hfill{\Box}$

Now we extend the above conclusion to the algebra of
$\{P_{kp}^{(n)}\}$, which is used in next section to prove our
main theorem.

\begin{theorem}\label{mlemma}
\begin{align}
P_{kp}^{(n)}=\bar{P}_{kp}^{(n)}+\sum_{i=1}^{n}\sum_{m_1,m_2,...,m_i\in\mathbb{N}}P_{-m_1p}P_{-m_2p}\cdot...\cdot
P_{-m_ip}(constant\cdot
\bar{P}_{(m_1+m_2+...+m_i+k)p}^{(n-i)}\\+constant\cdot\bar{P}_{(m_1+m_2+...+m_i+k)p}^{(n-1-i)}+...+constant\cdot
\bar{P}_{(m_1+m_2+...+m_i+k)p}^{(0)})\label{wp1}
\end{align}
where the constants only depend on $p$ and especially the
constant in front of $P_{-m_ip}\cdot \bar{P}_{(m_i+k)p}^{(n-1)}$
equals $n$.
\end{theorem}
\noindent{\sl{\bf Proof:}}  From the expression of $W_{kp}^{(n)}$
and $P_{kp}^{(n)}$, we know that one could be regard as a linear
representation of the other. That is,
\begin{equation}
W_{kp}^{(n)}=P_{kp}^{(n)}+constant\cdot
P_{kp}^{(n-1)}+...+constant\cdot P_{kp}^{(1)}.\label{pa1}
\end{equation}
and
\begin{equation}
P_{kp}^{(n)}=W_{kp}^{(n)}+constant\cdot
W_{kp}^{(n-1)}+...+constant\cdot W_{kp}^{(1)},\label{pa2}
\end{equation}
where the constants only depend on $kp$. Use \lemmaref{wa} and
substitute the expansion of $W_{kp}^{(n-i)}$ in $\{P_{-mp}\}$
into (\ref{pa2}). Then, we obtain
\begin{align}
P_{kp}^{(n)}&=\bar{W}_{kp}^{(n)}+constant\cdot
\bar{W}_{kp}^{(n-1)}+...+constant\cdot
\bar{W}_{kp}^{(2)}\\&+\sum_{i=1}^{n}\sum_{m_1,m_2,...,m_i\in\mathbb{N}}P_{-m_1p}P_{-m_2p}\cdot...\cdot
P_{-m_ip}(constant\cdot
\bar{W}_{(m_1+m_2+...+m_i+k)p}^{(n-i)}\\&+constant\cdot\bar{W}_{(m_1+m_2+...+m_i+k)p}^{(n-1-i)}+...+constant\cdot
\bar{W}_{(m_1+m_2+...+m_i+k)p}^{(0)}).\label{pa3}
\end{align}
Let $t_{mp}=0$ on both sides of (\ref{pa1}) and (\ref{pa2}) and
we obtain
\begin{equation}
\bar{W}_{kp}^{(n)}=\bar{P}_{kp}^{(n)}+constant\cdot
\bar{P}_{kp}^{(n-1)}+...+constant\cdot
\bar{P}_{kp}^{(1)},\label{pa4}
\end{equation}
and
\begin{equation}
\bar{P}_{kp}^{(n)}=\bar{W}_{kp}^{(n)}+constant\cdot
\bar{W}_{kp}^{(n-1)}+...+constant\cdot
\bar{W}_{kp}^{(1)}.\label{pa5}
\end{equation}
Substitute (\ref{pa4}) with $n=2,3,...,n-1$ and (\ref{pa5})  into
(\ref{pa3}). Then we obtain the expansion of (\ref{wp1}), in
which the constants are different from (\ref{pa3}), but they are
still depend only on $p$. Meanwhile, the item $P_{m_ip}\cdot
P_{(m_i+k)p}^{(n-1)}$ is generated only by the item of
$W_{kp}^{(n)}$. And the constant in front of it equals the
constant in front of $P_{-m_ip}\cdot W_{(m_i+k)p}^{(n-1)}$. So
the constant in front of $P_{-m_ip}\cdot P_{(m_i+k)p}^{(n-1)}$
also equals $n$.

$\hfill{\Box}$

In next section, we'll give the expansions of $P_{kp}^{(3)}$,
$P_{kp}^{(4)}$ and $P_{kp}^{(5)}$, which can be regard as
examples and verifications of \autoref{mlemma} with $n=3,4,5$.

These conclusions also hold for $\{\tilde{W}_{kp}^{(n)}\}$ and
$\{\tilde{P}_{kp}^{(n)}\}$, that is
\begin{corollary}
\begin{align}
P_{kp}^{(n)}=\tilde{P}_{kp}^{(n)}+\sum_{i=1}^{n}
\sum_{\substack{m_1,m_2,...,m_i;\\n_1,n_2,...n_i\in\mathbb{N}}}P_{-m_1p}P_{-m_2p}\cdot...\cdot
P_{-m_ip}(constant\cdot\tilde{P}_{(m_1+m_2+...+m_i+k)p}^{(n-i)}+\\
constant\cdot\tilde{P}_{(m_1+m_2+...+m_i+k)p}^{(n-1-i)}+...+constant\cdot
P_{(m_1+m_2+...+m_i+k)p}^{(0)})P_{n_1p}P_{n_2p}\cdot...\cdot
P_{n_ip}
\end{align}
where the constants only depend on $p$.

Similarly, there is the same connection between $W_{kp}^{(n)}$
and $\tilde{W}_{kp}^{(n)}$.
\end{corollary}
Based on this connection, we can calculate the commutation
relations of one algebra from those of another algebra. As an
example, we show the calculation for $n=2$. The conclusion
through the method is the same as those through a straightforward
calculation; but the calculation is much simpler. This can also
be regard as a verification of \autoref{mlemma} with $n=2$.
\begin{corollary}\label{coro3}
 $\{\frac{1}{2}\bar{P}_{kp}^{(2)}\}$ has the same commutation relations as
 $\{\frac{1}{2}P_{kp}^{(2)}\}$.
\end{corollary}
\noindent{\sl {\bf   Proof:}}  $\{\frac{1}{2}P_{kp}^{(2)}|
k=-1,0,1,\cdots\}$ is a subalgebra of the Virasoro algebra. Then
their commutation relations are
\begin{equation}
[\frac{1}{2}P_{np}^{(2)},
\frac{1}{2}P_{mp}^{(2)}]=\frac{1}{2}(np-mp)P_{np+mp}^{(2)}+\frac{1}{12}((np)^3-np)\delta_{np+mp,0},\label{vcr}
\end{equation}
where $m,n\in\{-1,0,1,\cdots\}$. By direct computation, we have
\begin{equation}
P_{kp}^{(2)}=\bar{P}_{kp}^{(2)}+2\cdot\sum_{m\in\mathbb{N}}P_{-mp}P_{(k+m)p}.
\end{equation}
Substitute it into (\ref{vcr}) and note that $\bar{P}_{kp}^{(2)}$
don't include the variables of $\{t_{mp}\}$ so that they are
commutable with $P_{kp}$. So we have
\begin{align}
&\frac{1}{2}(np-mp)\bar{P}_{np+mp}^{(2)}+(n-m)p\cdot\sum_{l_3\in\mathbb{N}}P_{-l_3p}P_{(l_3+n+m)p}+\frac{1}{12}((np)^3-np)\delta_{np+mp,0}\\
=&[\frac{1}{2}\bar{P}_{np}^{(2)}+\sum_{l_1\in\mathbb{N}}P_{-l_1p}P_{(l_1+n)p},
\frac{1}{2}\bar{P}_{mp}^{(2)}+\sum_{l_2\in\mathbb{N}}P_{-l_2p}P_{(l_1+m)p}]\\
=&[\frac{1}{2}\bar{P}_{np}^{(2)},\frac{1}{2}\bar{P}_{mp}^{(2)}]+[\sum_{l_1\in\mathbb{N}}P_{-l_1p}P_{(l_1+n)p}
\sum_{l_2\in\mathbb{N}}P_{-l_2p}P_{(l_1+m)p}]\\
=&[\frac{1}{2}\bar{P}_{np}^{(2)},\frac{1}{2}\bar{P}_{mp}^{(2)}]+\sum_{n+l_1=l_2}P_{-l_1p}\cdot
(n+l_1)p\cdot P_{(m+l_2)p}-\sum_{m+l_2=l_1}P_{-l_2p}\cdot
(m+l_2)p\cdot P_{(n+l_1)p}\\
=&[\frac{1}{2}\bar{P}_{np}^{(2)},\frac{1}{2}\bar{P}_{mp}^{(2)}]+\sum_{l_1\in\mathbb{N}}P_{-l_1p}\cdot
(n-m)p\cdot P_{(m+n+l_1)p}
\end{align}
Cancel the same item on both sides and we have
\begin{align}
[\frac{1}{2}P_{np}|_{t_{lp}=0}^{(2)},
\frac{1}{2}P_{mp}|_{t_{lp}=0}^{(2)}]=&[\frac{1}{2}P_{np}^{(2)},
\frac{1}{2}P_{mp}^{(2)}]|_{t_{lp}=0}\\
=&\frac{1}{2}(np-mp)P_{np+mp}^{(2)}+\frac{1}{12}((np)^3-np)\delta_{np+mp,0};
\quad l\in\mathbb{N}
\end{align}
which is the same as $\{\frac{1}{2}P_{kp}^{(2)}\}$ $\hfill{\Box}$

\section*{\S 4 Precise constraints on a associated tau function }
\setcounter{section}{4} \setcounter{equation}{0}

\quad In this section, based on \theoremref{mlemma} we propose a
new and computable method which can determine the values of
unknown constants in higher order constraints on a $\tau(t)$
function which satisfies both the p-reduced KP hierarchy and the
sting equation. Consequently, with these values we can obtain a
precise algebra that the higher constraints compose. Meanwhile,
the algebra includes the Virasoro algebra as its subalgebra. So,
the conclusion in \cite{dvv,fkn}, which is that a $\tau$ function
under the constraints of 2-reduced KP and the string equation is
a vacuum vector of the Virasoro algebra, is also included in our
conclusion.

Now we first give a preliminary conclusion which shows how the
unknown constants in constraints on a corresponding $\tau$
function generates. Similar result was obtained through other
method. Here we use a simple way to obtain it.
\begin{lemma}\label{coro2}
When $k=-1,0,1...; i\in\mathbb{N}$
\begin{equation}
\bar{P}_{kp}^{(i)}\cdot\tau(t)=c_k^{(i)}\cdot\tau(t),
\end{equation} where each $c_k^{(i)}$ is a constant.
\end{lemma}
\noindent{\sl {\bf   Proof:}} From (\ref{69}), we obtain
\begin{equation}
W_{kp}^{(j+1)}\cdot\tau(t)=constant\cdot\tau(t)\quad
\text{when}\quad k=0,1,2,\cdots; \quad j=1,2,\cdots.\label{p6}
\end{equation}
Note a fact that
\begin{equation}
-\prod_{r=0}^{j-1}(\frac{p-1}{2}-r)(G(z)-1)(\dfrac{P_{-p}\cdot\tau(t)}{\tau(t)})=\prod_{r=0}^{j-1}(\frac{p-1}{2}-r)z^{-p}.\label{p3}
\end{equation}
Add (\ref{p3}) to (\ref{p4}) and we obtain
\begin{equation}
(G(z)-1)\dfrac{\big(\dfrac{W_{kp}^{{(j+1)}}}{j+1}-\prod_{r=0}^{j-1}(\dfrac{p-1}{2}-r)t_p\big)\cdot\tau(t)}{\tau(t)}
=0\quad \text{when}\quad k=-1; \quad j\in\mathbb{N}.
\end{equation}
That is
\begin{equation}
\big(W_{kp}^{{(j+1)}}-(j+1)\prod_{r=0}^{j-1}(\dfrac{p-1}{2}-r)t_p\big)\cdot\tau(t)=constant\cdot\tau(t)\quad
\text{when}\quad k=-1; \quad j\in\mathbb{N}.\label{p77}
\end{equation}
Consider (\ref{p6})and (\ref{p77}) together. Let $t_{mp}=0$ on
both sides of the two identities. We could cancel the items that
include variables of $\{t_{mp}\}$. Since the $\tau(t)$ of a
p-reduced KP is independent on the variables of $\{t_{mp}\}$, we
obtain
\begin{equation}
\bar{W}_{kp}^{(j+1)}\cdot\tau(t)=constant\cdot\tau(t)\quad
\text{when}\quad k=-1,0,1,2,\cdots; \quad
j\in\mathbb{N}.\label{p7}
\end{equation}
Meanwhile, we have
\begin{equation}
P_{kp}^{(n)}=W_{kp}^{(n)}+constant\cdot
W_{kp}^{(n-1)}+...+constant\cdot W_{kp}^{(1)}.
\end{equation}
Let $t_{mp}=0$ on both sides, and we have
\begin{equation}
\bar{P}_{kp}^{(n)}=\bar{W}_{kp}^{(n)}+constant\cdot
\bar{W}_{kp}^{(n-1)}+...+constant\cdot \bar{W}_{kp}^{(1)}.
\end{equation}
So, together with \eqref{p7} and that the $\tau$ function is
independent on $t_{mp}$, we obtain
\begin{equation}
\bar{P}_{kp}^{(n)}\cdot\tau(t)=constant\cdot\tau(t)\quad k=-1,0,1...; i\in\mathbb{N}
\end{equation}

$\hfill{\Box}$

From the above lemma, we know that there are a lot of unknown
constants, $c_k^{(i)}$, in the constraints. Now, we propose a
method to determine the values of these constants and obtain the
precise constraints.

\begin{theorem}\label{constants}
The constants $c_k^{(i)},k=0,1,2...,i\in\mathbb{N}$ can be
determined by a recursive process step by step, in which
\begin{equation}
c_k^{(i)}=0,k=1,2,3....
\end{equation}
and
\begin{equation}
c_0^{(1)}(p)=0,\quad c_0^{(2)}(p)=-\frac{1}{12}(p^2-1),\quad
c_0^{(3)}(p)=0,\quad
c_0^{(4)}(p)=-(\frac{7}{240}p^2-\frac{1}{80})(p^2-1)\cdots
\end{equation}
\end{theorem}

\noindent{\sl {\bf   Proof:}} This method is essentially  a
recursive process. Here we use the second mathematical reduction
on $i$ to prove it. Start from $i=2$. Let $k=-1$ and $j=2$ in
(\ref{p4}), and we have
\begin{equation}
(G(z)-1)\frac{\frac{1}{3}(P_{-p}^{(3)}-\frac{3}{2}(-p+2)P_{-p}^{(2)}+(-p+1)(-p+2)P_{-p})\cdot\tau(t)}{\tau(t)}=-\frac{p-3}{2}\cdot\frac{p-1}{2}z^{-p}.\label{p8}
\end{equation}
Applying \lemmaref{mlemma} with $n=2,3$ and by a straightforward
computation, we have
\begin{equation}
P_{-p}^{(3)}\cdot\tau(t)=\bar{P}_{-p}^{(3)}\cdot\tau(t)+\big(\sum_{m=1}^{\infty}
3\cdot
P_{-mp}\cdot\bar{P}^{(2)}_{(m-1)p}\big)\cdot\tau(t)+\big(\sum_{n,l\in\mathbb{N}}3\cdot
P_{-np}\cdot P_{-lp}\cdot
P_{(n+l-1)p}\big)\cdot\tau(t).\label{p9}
\end{equation}
Here the third item in the right-hand side of the above identity
equals $0$ since $n+l-1\geq0$ and $\tau(t)$ is independent on
$t_{mp}$; using \lemmaref{coro2}, the fist two items in the
right-hand side equal
$c_{-1}^{(3)}\cdot\tau(t)+\big(\sum_{m=1}^{\infty} 3\cdot
P_{-mp}\cdot c^{(2)}_{(m-1)}\big)\cdot\tau(t)$. As for
$P_{-p}^{(2)}$, the conclusion is similar. And we have
\begin{equation}
P_{-p}^{(2)}\cdot\tau(t)=\bar{P}_{-p}^{(2)}\cdot\tau(t)+\sum_{n=1}^{\infty}2\cdot
P_{-np}\cdot
P_{(n-1)p}\tau(t)=c_{-1}^{(2)}\cdot\tau(t)\label{p10}
\end{equation}
Substitute (\ref{p9}) and (\ref{p10}) into (\ref{p8}) and we have
\begin{equation}
(G(z)-1)\big(\sum_{m=1}^{\infty}
 P_{-mp}\cdot c^{(2)}_{(m-1)p}+\frac{1}{3}(-p+1)(-p+2)P_{-p}\big)=-\frac{p-3}{2}\cdot\frac{p-1}{2}z^{-p}.
\end{equation}
By straightforward computation, it is
\begin{equation}
\sum_{m=1}^{\infty}
 z^{-mp}\cdot c^{(2)}_{(m-1)p}+\frac{1}{3}(-p+1)(-p+2)z^{-p}=\frac{p-3}{2}\cdot\frac{p-1}{2}z^{-p}.
\end{equation}
Comparing the coefficient of $z^{-mp}$ on both sides, we obtain
\begin{eqnarray}
\begin{cases}
c_0^{(2)}=-\dfrac{1}{12}(p^2-1),\\\\
c_k^{(2)}=0;\quad k=1,2,3,\cdots.
\end{cases}
\end{eqnarray}
So the theorem holds when $i=2$.

Assume the theorem is holds for $i\leq n-2$, that is
 $c_k^{(i)}=0$ where $k\in\mathbb{N}$ and $c_0^{(i)}$ are already
determined.

Now we prove the theorem holds for $i=n-1$. From \eqref{p4} with
$j=n-1$, we have
\begin{equation}
(G(z)-1)\dfrac{\dfrac{W_{-p}^{{(n)}}}{n}\cdot\tau(t)}{\tau(t)}=-\prod_{r=0}^{n-2}(\frac{p-1}{2}-r)z^{-p},\label{temp1}
\end{equation}
in which
\begin{equation}
W_{-p}^{(n)}\cdot\tau(t)=(P_{-p}^{(n)}+constant\cdot P_{-p}^{(n-1)}+...+constant\cdot
P_{-p}^{(1)})\cdot\tau(t).
\end{equation}
Applying \theoremref{mlemma} with $k=-1$. Because $\sum_i -m_i\neq-1$
when $i\geq2$ and $P_{-p}^{(j)}$ doesn't include a single item of $P_{-mp}^{(1)}$ when $j\geq2$,
we can cancel some constants in the expansion of $P_{-p}^{(j)}$.
Then each $P_{-p}^{(j)}$ with $j\geq2$ in the above identity can be expanded as
\begin{align}
P_{-p}^{(j)}&=\bar{P}_{-p}^{(j)}+j\cdot\sum_{m=1}^{\infty}P_{-mp}(\sum_{k=1}^{j-1} \bar{P}_{(m-1)p}^{(k)})\\
&+\sum_{m_1,m_2\in\mathbb{N}}P_{-m_1p}P_{-m_2p}(\sum_{k=1}^{j-2}constant\cdot \bar{P}_{(m_1+m_2-1)p}^{(k)})\\&+...  ...\\
&+\sum_{m_1,m_2,...m_{j-1}}P_{-m_1p}P_{-m_2p}\cdot...\cdot P_{-m_{j-1}p}(constant\cdot \bar{P}_{(m_1+m_2+...m_{j-1}-1)p}^{(1)})
\end{align}
Substitute the above into (\ref{temp1}) and we obtain
\begin{equation}
(G(z)-1)\frac{A}{n\tau(t)}
=-\prod_{r=0}^{n-2}(\frac{p-1}{2}-r)z^{-p}
\end{equation}
where
\begin{align}
A&=\sum\limits_{m\in\mathbb{N}} P_{-mp}
(n\cdot\sum\limits_{k=1}^{n-1}c_{m-1}^{(k)})+\sum_{i=2}^{n-1}\sum\limits_{m_1,m_2,...m_i\in\mathbb{N}}\prod_{l=1}^{i}
P_{-m_lp}(\sum\limits_{k=1}^{n-i}constant\cdot c_{m_1+m_2+...+m_i-1}^{(k)})\\
&+constant\cdot(\sum_{i=1}^{n-2}\sum\limits_{m_1,m_2,...m_i\in\mathbb{N}}\prod_{l=1}^{i}
P_{-m_lp}(\sum\limits_{k=1}^{n-1-i}constant\cdot
c_{m_1+m_2+...+m_i-1}^{(k)}))\\
&+\cdots \cdots\\
&+constant\cdot P_{-p}^{(1)}
\end{align}
Substitute the results of $i\leq n-2$ into the above, we obtain
\begin{align}
(G(z)-1)&\frac{
\begin{pmatrix}
\sum\limits_{m\in\mathbb{N}}P_{-mp}\cdot
c_{m-1}^{(n-1)}+\sum\limits_{l=2}^{n-1}constant\cdot P_{-p}\cdot
c_0^{k-l}+constant\cdot P_{-p}
\end{pmatrix}\cdot\tau(t)}{\tau(t)}\cdot\notag\\ =&-\prod_{r=0}^{n-2}(\frac{p-1}{2}-r)z^{-p}
\end{align}
Comparing the coefficients of $z^{mp}$ on both sides, we obtain
\begin{eqnarray}
\begin{cases}
c_0^{(n-1)}+\sum\limits_{l=2}^{n-1}constant\cdot
c_0^{k-l}+constant=-\prod\limits_{r=0}^{n-2}(\frac{p-1}{2}-r),\\\\
c_k^{(n-1)}=0;\quad k=1,2,3,\cdots.
\end{cases}
\end{eqnarray}
There is only one unknown variable, $c_0^{(n-1)}$, in the first
equation. Then we can determined it from that. So $c_{k}^{n-1}=0$
where $k\in\mathbb{N}$ and $c_0^{(n-1)}$ is also determined, that
is, the theorem holds for $i=n-1$. According to the second
mathematical reduction, the theorem holds. The proof also give us
a recursive method which can calculate the $c_k^{(i)}$ step by
step.

$\hfill{\Box}$

From the above proof, we know that the method is essentially a
recursive process, that is, we could determine the constants step
by step. When $p=2$, $c_0^{(2)}=-\frac{1}{12}(p^2-1)$ through
this method, which is the same as $c_0^{(2)}$ obtained through a
direct computation of commutation relations of Virasoro algebra
in \autoref{coro3}. Moreover, it is usually that assign $0$ to
all $c_k^{(i)}$, but here, using this method, we know that all of
them do not equal $0$. The constants which do not equal $0$ are
close related to the centers of the algebra that the constraints
constitute. As for $c_{-1}^{(i)}$, the are unknown; but
$c_{-1}^{(2)}=0$, which can obtained by the commutation relations
of the Virasoro algebra.

Now, with these values of the constants, we could obtain the
precise algebra of the constraints. Note that the above theorem
is still holds if we substitute $\{\tilde{P_{kp}^{(i)}}\}$ for
$\{\bar{P_{kp}^{(i)}}\}$. So, we have
\begin{corollary}
If $L$ satisfies the p-reduced KP hierarchy and
$\frac{p-1}{2}L^{-p}=(ML^{-p+1})_-$, then $L$ satisfies the
String equation and the $\tau$ function of $L$ is a vacuum vector
of the algebra $\bar{P}$ or $\tilde{P}$ where
\begin{equation}
P=\{\bar{P}^{(i)}_{-p}-c_{-1}^{(i)},\bar{P}^{(i)}_0-c_0^{(i)},
\bar{P}^{(i)}_{kp}|i, k\in\mathbb{N}\}.
\end{equation}
and
\begin{equation}
P=\{\tilde{P}^{(i)}_{-p}-c_{-1}^{(i)},\tilde{P}^{(i)}_0-c_0^{(i)},
\tilde{P}^{(i)}_{kp}|i, k\in\mathbb{N}\}.
\end{equation}
That is
\begin{equation}
\bar{P}\cdot\tau(t)=0 \quad\text{and}\quad
\tilde{P}\cdot\tau(t)=0.
\end{equation}
\end{corollary}
\begin{corollary}
The algebra $P$ includes a Virasoro algebra with no center as its
subalgebra, that is, $\{\bar{P}_{np}^{(2)}-c_n^{(2)}\}$.
\end{corollary}
\noindent{\sl {\bf   Proof:}} Let
\begin{equation}
\dfrac{1}{2}(\bar{P}_{np}^{(2)}-c_n^{(2)})=L_n,n=-1,0,1,2,...
\end{equation}
Applying \autoref{coro3}, we have
\begin{equation}
[L_n, L_m]=(n-m)L_{n+m}.
\end{equation}

$\hfill{\Box}$

So, the conclusion in \cite{dvv,fkn}, which is that a $\tau$
function under the constraints of 2-reduced KP and the string
equation is a vacuum vector of the Virasoro algebra, is also
included in our conclusion.in \cite{dvv,fkn}. That is, we extend
the sufficiency of the above conclusion from 2-reduced KP and the
Virasoro algebra to an arbitrary p-reduced KP and a algebra of
$\bar{P}$ or $\tilde{P}$.

Now we show the detailed calculation when $i=4$ and $5$, which
together with the case of $i=3$ can be used as examples of
\autoref{constants} and \autoref{mlemma}.

Consider the case of $i=j+1=4$ in (\ref{p4}). We have
\begin{align}
&(G(z)-1)\frac{\frac{1}{4}(P_{-p}^{(4)}-(6-2p)P_{-p}^{(3)}+(2p^2-9p+11)P_{-p}^{(2)}-(1-p)(2-p)(3-p)P_{-p})\cdot\tau(t)}{\tau(t)}\notag\\
&=-\frac{p-5}{2}\cdot\frac{p-3}{2}\cdot\frac{p-1}{2}z^{-p}\label{610}
\end{align}
By a straightforward computation,
\begin{align}
P_{-p}^{(4)}&=\bar P_{-p}^{(4)}+4\cdot
\sum_{m\in\mathbb{N}}P_{-mp}\cdot\bar{P}_{mp-p}^{(3)}+\sum_{m_1,m_2\in\mathbb{N}}6\cdot
P_{-m_1p}\cdot P_{-m_2p}\cdot
P_{(m_1+m_2-1)p}^{(2)}\\&+\sum_{m\in\mathbb{N}}2\cdot
mp(mp-p)P_{-mp}P_{mp-p}+\sum_{m_1,m_2,m_3\in\mathbb{N}}4\cdot
P_{-m_1p}P_{-m_2p}P_{-m_3p}P_{(m_1+m_2+m_3-1)p}.\label{p11}
\end{align}
Substitute (\ref{p9}) and (\ref{p11}) into (\ref{610}) and we
have
\begin{align}
(G(z)-1)&\frac{\sum\limits_{m\in\mathbb{N}} P_{-mp}\cdot
c_{m-1}^{(3)}+\frac{3}{2}(p-3)\sum\limits_{m\in\mathbb{N}}
\cdot P_{-mp}\cdot c_{m-1}^{(2)}+\frac{1}{4}(p-1)(p-2)(p-3)P_{-p}\cdot\tau(t)}{\tau(t)}\notag\\
&=-\frac{p-5}{2}\cdot\frac{p-3}{2}\cdot\frac{p-1}{2}z^{-p}.\label{612}
\end{align}
Comparing the coefficient of $z^{-mp}$ on both sides, we obtain
\begin{eqnarray}
\begin{cases}
c_0^{(3)}-\dfrac{(p^2-1)(p-3)}{8}+\dfrac{(p-1)(p-2)(p-3)}{4}=\dfrac{(p-1)(p-3)(p-5)}{8},\\\\
c_{m-1}^{(3)}=0,\quad m=2,3,4,\cdots.
\end{cases}
\end{eqnarray}
Solving them, we have
\begin{eqnarray}
\begin{cases}
c_0^{(3)}=0\\\\
c_{m-1}^{(3)}=0,\quad m=2,3,4,\cdots.
\end{cases}
\end{eqnarray}
Consider the case of $i=j+1=5$ in (\ref{p4}). we have
\begin{align}
(G(z)-1)&\frac{\frac{1}{5}\cdot
\begin{pmatrix}
(P_{-p}^{(5)}-\frac{5}{2}(4-p)P_{-p}^{(4)}+(\frac{10}{3}p^2-20p+35)P_{-p}^{(3)}
-\\(-\frac{5}{2}p^3+20p^2-\frac{105}{2}p+50)P_{-p}^{(2)}+(1-p)(2-p)(3-p)(4-p)P_n)
\end{pmatrix}\cdot\tau(t)}{\tau(t)}\cdot\notag\\ =&-\prod_{r=0}^{3}(\frac{p-1}{2}-r)z^{-p} \label{p15}
\end{align}
In order to obtain the expansion of $P_{-p}^{(5)}$ in $\{t_{mp}\}$, we first calculate the items in $P_{-p}^{(5)}$ which include only one variable in $\{t_{mp}\}$. Note that
\begin{equation}
2\sum_{i+j=kp}:iP_iP_j:=\sum_{i+j=kp}:iP_iP_j:+\sum_{i+j=kp}j:P_jP_i:\\
=\sum_{i+j=kp}kp:P_iP_j:.
\end{equation}
Let $\{t_{mp}|m\in\mathbb{N}\}=0$ on both sides, and we have
\begin{equation}
2\sum_{\substack{i+j=kp\\ i,j\neq-mp}}:iP_iP_j:=\sum_{\substack{i+j=kp\\ i,j\neq-mp}}:iP_iP_j:+\sum_{\substack{i+j=kp\\ i,j\neq-mp}}:jP_jP_i:\\
=kp\cdot\bar{P}_{kp}^{(2)}.
\end{equation}
So the items which include only one variables in $t_{mp}$ in $P_{-p}^{(5)}$ are
\begin{align}
&\sum_{m\in\mathbb{N}}5\cdot
P_{-mp}\sum_{\substack{i+j+k+l=mp-p\\
i,j,k,l\neq-mp}}:P_iP_jP_kP_l:-\sum_{m\in\mathbb{N}}5\cdot
P_{-mp}\sum_{\substack{i+j=mp-p\\ i,j\neq-mp}}:ijP_iP_j:\notag\\
&-\sum_{m\in\mathbb{N}}5\cdot(-mp)\cdot
P_{-mp}\sum_{\substack{i+j=mp-p\\
i,j\neq-mp}}:jP_iP_j:-\sum_{m\in\mathbb{M}}5\cdot(-mp)\cdot
P_{-mp}\sum_{\substack{i+j=mp-p\\ i,j\neq-mp}}:i P_iP_k:\notag\\
=&\sum_{m\in\mathbb{N}}5\cdot
P_{-mp}\sum_{\substack{i+j+k+l=mp-p\\
i,j,k,l\neq-mp}}:P_iP_jP_kP_l:-\sum_{m\in\mathbb{N}}5\cdot
P_{-mp}\sum_{\substack{i+j=mp-p\\ i,j\neq-mp}}:ijP_iP_j:\notag\\
&+\sum_{m\in\mathbb{N}}5mp\cdot P_{-mp}\sum_{\substack{i+j=mp-p\\ i,j\neq-mp}}(mp-p):P_iP_j:\notag\\
=&\sum_{m\in\mathbb{N}}5\cdot P_{-mp}\cdot
\bar{P}_{mp-p}^{(4)}+\sum_{m\in\mathbb{N}}5mp(mp-p)P_{-mp}\bar{P}_{mp-p}^{(2)}.
\end{align}
Substitute the above into the expansion. Then it can be written
as
\begin{align}
P_{-p}^{(5)}&=\bar{P}_{-p}^{(5)}+\sum_{m\in\mathbb{N}}5\cdot
P_{-mp}\cdot
\bar{P}_{mp-p}^{(4)}+\sum_{m\in\mathbb{N}}5mp(mp-p)P_{-mp}\bar{P}_{mp-p}^{(2)}\\
&+\sum_{m_1,m_2\in\mathbb{N}}10\cdot P_{-m_1p}\cdot P_{-m_2p}\cdot \bar{P}_{(m_1+m_2-1)p}^{(3)}\\
&+\sum_{m_1,m_2\in\mathbb{N}}5\cdot(m_1\cdot m_2+m_1(m_1+m_2-p)+m_2(m_1+m_2-p)P_{-m_1p}P_{-m_2p}P_{(m_1+m_2-1)p}\\
&+\sum_{m_1,m_2,m_3\in\mathbb{N}}10\cdot P_{-m_1p}P_{-m_2p}P_{-m_3p}\bar{P}_{(m_1+m_2+m_3-1)p}^{(2)}\\
&+\sum_{m_1,m_2,m_3,m_4\in\mathbb{N}}5\cdot
P_{-m_1p}P_{-m_2p}P_{-m_3p}P_{-m_4p}P_{(m_1+m_2+m_3+m_4-1)p}.
\end{align}
Substitute the above expansion into (\ref{p15}). Then we obtain
\begin{align}
(G(z)-1)&\frac{\begin{pmatrix} \sum_{m\in\mathbb{N}} P_{-mp}\cdot
c_{m-1}^{(4)}+\sum_{m\in\mathbb{N}}m(m-1)p^2P_{-mp}\cdot
c_{m-1}^{(2)}-(8-2p)\sum_{m\in\mathbb{N}}P_{-mp}\cdot
c_{m-1}^{(3)}\\+(2p^2-12p+21)\sum_{m\in\mathbb{N}}P_{-mp}\cdot\cdot
c_{m-1}^{(2)} +\frac{1}{5}(1-p)(2-p)(3-p)(4-p)P_{-p}
\end{pmatrix}
}{\tau(t)}\notag\\
&=-\frac{p-7}{2}\frac{p-5}{2}\cdot\frac{p-3}{2}\cdot\frac{p-1}{2}z^{-p}.\label{612}
\end{align}
Comparing the coefficient of $z^{-mp}$ on both sides, we obtain
\begin{eqnarray}
\begin{cases}
c_0^{(4)}-(\frac{1}{6}p^2-p+\frac{7}{4})(p^2-1)+\frac{1}{5}(p-1)(p-2)(p-3)(p-4)
=\frac{p-1}{2}\cdot\frac{p-3}{2}\cdot\frac{p-5}{2}\cdot\frac{p-7}{2}.,\\\\
c_{m-1}^{(3)}=0,\quad m=2,3,4,\cdots.
\end{cases}
\end{eqnarray}
Solving them, we have
\begin{eqnarray}
\begin{cases}
c_0^{(4)}=(\frac{7}{240}p^2-\frac{1}{80})(p^2-1),\\\\
c_{m-1}^{(3)}=0,\quad
m=2,3,4,\cdots.
\end{cases}
\end{eqnarray}

$\hfill{\Box}$

\section*{\S 5 Conclusions}
\setcounter{section}{5} \setcounter{equation}{0}

\quad  In this study, for an arbitrary $p$, we propose a new
method which can determine the values of unknown constants in
constraints on a tau function  which satisfies both the p-reduced
KP hierarchy and the sting equation. It is a recursive process
and through it we could directly calculate the constants step by
step. By this method, we know that, unlike what people usually
think of, all of them do not equal $0$. When $p=2$, our
conclusion is the same as the current conclusion, that is, the
constants determined through our method being the same as those
determined through commutation relations of Virasoro. With these
values we obtain the precise algebra that the constraints
compose. Meanwhile, the algebra includes the Virasoro algebra as
its subalgebra. So, the conclusion in \cite{dvv,fkn}, which is
that a $\tau$ function under the constraints of 2-reduced KP and
the string equation is a vacuum vector of the Virasoro algebra,
is also included in our conclusion. That is, we also extend the
sufficiency of the above conclusion from 2-reduced KP and the
Virasoro algebra to an arbitrary p-reduced KP and a algebra of
$\bar{P}$ or $\tilde{P}$.

In this process we also obtain the connection between the
$W_{1+\infty}$ algebra which includes the redundant variables of
$\{t_{mp}\}$ and the algebra of
$\bar{W}=\{W_{n}^{(m)}|_{t_{mp}=0}\}$ which doesn't include
$\{t_{mp}\}$. Through this connection, we could obtain
commutation relations of one algebra from those of the other
algebra. And the calculation is much simpler than a
straightforward calculation.

This method is a general approach which is feasible to other
integrable systems similar to the KP system, such as dKP
hierarchy, BKP hierarchy, qKP hierarchy etc. In the near further.
we will try them. And we will study further the algebraic structure
of the algebra $\bar{P}$. \\



\end{document}